\documentclass[10 pt,twocolumn,notitlepage]{article}
\usepackage{amsmath}
\usepackage{amssymb}
\usepackage{graphicx}
\usepackage{titling}
\usepackage{subcaption}
\usepackage[toc,page]{appendix}
\usepackage{lipsum}
\usepackage{float}
\usepackage{titling}
\usepackage[left=2cm,right=2cm,top=2cm,bottom=2cm]{geometry}
\usepackage{wrapfig}
\usepackage[capitalize,noabbrev]{cleveref}
 \usepackage{algorithm}
 \usepackage{algorithmic}

\title{High Accuracy Uncertainty-Aware Interatomic Force Modeling with Equivariant Bayesian Neural Networks}

\author{
Tim Rensmeyer, Benjamin Craig, Denis Kramer, Oliver Niggemann
}
\date{}
\begin{document}
\maketitle

\thispagestyle{empty}
%\twocolumn[

%\vskip 0.3in
%]

\begin{abstract}
Even though Bayesian neural networks offer a promising framework for modeling uncertainty, active learning and incorporating prior physical knowledge, few applications of them can be found in the context of interatomic force modeling.
One of the main challenges in their application to learning interatomic forces is the lack of suitable Monte Carlo Markov chain sampling algorithms for the posterior density, as the commonly used algorithms do not converge in a practical amount of time for many of the state-of-the-art architectures.
As a response to this challenge, we introduce a new Monte Carlo Markov chain sampling algorithm in this paper which can circumvent the problems of the existing sampling methods.
In addition, we introduce a new stochastic neural network model based on the NequIP architecture and demonstrate that, when combined with our novel sampling algorithm, we obtain predictions with state-of-the-art accuracy as well as a good measure of uncertainty.
\end{abstract}

\section{Introduction}
Despite the fact that the laws of quantum mechanics, which underly chemistry, were discovered almost a century ago, their application for the {\it ab initio} prediction of many chemical phenomena remains a formidable task \cite{Atkins,MD_NNP}.
This is in large parts a result of the sheer computational complexity involved in solving these equations numerically as well as the enormous phase space that needs to be considered\cite{Atkins,MD_NNP}.\\
Molecular dynamics (MD), where the time evolution of molecular systems is investigated, is particularly affected by this challenge.
While computational methods such as Density Functional Theory (DFT) have been developed that can calculate interatomic forces with very high accuracy, these methods are in general computationally expensive and scale badly with growing system sizes.
Thus, it remains very difficult to model larger systems with high accuracy and a large enough time horizon.\\
To circumvent these difficulties, the learning of interatomic forces via machine learning models such as neural networks has become a highly active area of research \cite{Survey_NNP}. 
However, the cost of generating training data from {\it ab initio} simulations prevents the generation of large training sets commonly required to construct sufficiently predictive neural networks.
Recent innovations in neural network designs have already made it possible to learn highly accurate interatomic force fields for a given compound with limited data by incorporating hard constraints in the form of symmetry properties and energy conservation into the model architecture \cite{GemNet,SpookyNet,PaiNN,NequIP,NewtonNet,UNiTE}.
However, training models to have chemical accuracy for entire classes of compounds still requires large amounts of data \cite{GemNet,SpookyNet,NewtonNet,UNiTE}, owing in large part to the vastness of the space of possible atomic configurations.
The creation of predictive and transferable models is, therefore, still a challenging task.\\
Additionally, the existing state-of-the-art models generally lack a measure of uncertainty in their prediction \cite{Survey_NNP,Survey_GNNP},
 which makes it impossible to tell which of the predictions are reliable. 
This limits the applicability of active learning strategies, where uncertainty can be used to select useful training data from the large configuration space more efficiently \cite{AL3,AL2,AL4,AL1,MCMC_BNNP} by only labeling outlier configurations where the model predicts a high likelihood of a large error.\\
Further, for a deployed model detected outliers can either be recomputed on-the-fly in DFT to ensure the accuracy of a neural network-based MD trajectory or used to retroactively discard trajectories that are likely to be physically inaccurate.
All of these factors hinder their extension towards more general models, required for the prediction and optimization of complex (dynamic) properties of materials, where a vast amount of diverse atomic configurations need to be screened with an overall high accuracy and a good measure of uncertainty. The latter can be used to decide which of the samples that are predicted, should undergo additional investigation through {\it ab initio} calculations or experiments and which are likely to be false positives. An uncertainty measure is also attractive to ensure high accuracy of modeling rare events that often govern longer time-scale dynamics, but are seriously under-represented in MD trajectories \cite{AL1}. Configurations close to transition states are exponentially less likely to be sampled relative to near ground-state configurations. But these states often have peculiar electronic configurations and bonding characteristics that do not feature in low energy configurations. Hence, an uncertainty measure would allow on-the-fly identification of configurations that require additional learning to bias active learning strategies towards relevant "new chemistry" such as transition states.
\\
Furthermore, many existing state-of-the-art models are currently optimized in a somewhat intricate way on smaller datasets to improve performance, where artificial noise is introduced to the optimization process via a large learning rate and small batch size \cite{GemNet,SpookyNet,PaiNN,NequIP,UNiTE}. While this works well empirically, there is no established theoretical foundation for the utility of such optimization and consequently, there is no guide in choosing appropriate batch sizes and learning rates except trial and error.  
\\
As a response to these challenges, we investigate here the viability of combining Bayesian Neural Networks (BNNs) with state-of-the-art neural networks for modeling interatomic forces, as BNNs have a robust measure of uncertainty and an inherent principled way of adding noise to the optimization process via Monte Carlo Markov Chain (MCMC) methods. Further, they offer a good framework for the incorporation and data-based updating of prior beliefs, such as approximate physics-guided interatomic force-fields, which may be utilized in the future to further improve data efficiency and generalization.
\\
However, we found that current off-the-shelf MCMC methods are unsuited for dealing with the vastly different gradient scales that different parameter groups exhibit in modern models. This results in an unpractically slow traversal through the parameter space to the point where no convergence is achievable in reasonable amounts of time, which explains their current lack of use. 
Hence, a new MCMC algorithm is needed for high-quality Bayesian uncertainty quantification with neural network-based interatomic force models.

The contributions of this paper are the following:
\begin{itemize}
\item We develop a new stochastic neural network model based on the NequIP architecture \cite{NequIP} for uncertainty-aware interatomic force modeling with state-of-the-art accuracy.
\item We provide a novel Bayesian MCMC algorithm that produces high-accuracy models with high-quality uncertainty estimates in a practical amount of time without relying on finetuned batch sizes and learning rates.
\item We show that the current optimization methods based on high learning rates and small batch sizes already mimic Bayesian MCMC methods to a degree.
\item We discuss shortcomings as well as possible further improvements and extensions of the proposed neural network model and sampling algorithm.
\end{itemize}
Notation commonly used in this paper is summarized in Table 1.
\begin{table}[H]

%\caption{Commonly used Notation}
\begin{centering}
\caption{Commonly used notation}
\vskip 0.1in
\includegraphics[scale=0.85]{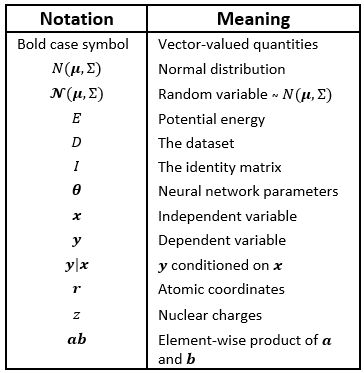}

\end{centering}
\end{table}

\subsection{Related Work}
\begin{figure*}
\begin{centering}
\includegraphics[scale=0.5]{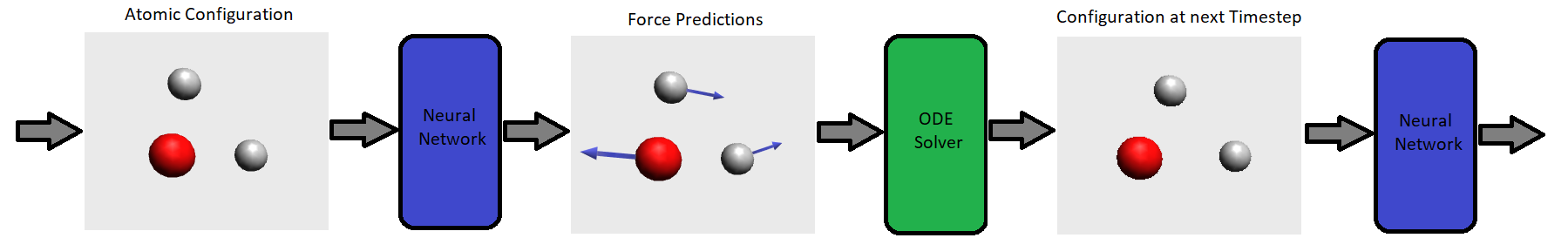}
\caption{An illustration of a molecular dynamics workflow with neural network predictions for the forces.}
\end{centering}
\end{figure*}
While BNNs to date have found very few applications in this context, some use cases exist in the literature \cite{MCMC_BNNP,BNNP_DO1,BNNP_DO2,MCMCfailure}.
%However, all of these instances were limited to very small fully connected neural networks while we utilize a state-of-the-art graph neural network based architecture.
However, in \cite{BNNP_DO1,BNNP_DO2} Monte Carlo dropout was used for Bayesian inference. While this method can in some instances be interpreted as Bayesian variational inference \cite{BNNP_DO}, it yields a poor approximation of the true posterior distribution even for variational inference-based methods \cite{BNNP_DO} and typically results in much poorer uncertainty quantification than MCMC methods \cite{BNN_quality1}.
While a much more accurate Bayesian algorithm was used in \cite{MCMC_BNNP}, the use case was very limited in scope to diatomic systems.
Further, all of the above instances were limited to very small fully connected neural networks while we utilize a state-of-the-art graph neural network-based architecture.
We only found one attempt to reconcile graph neural network-based architectures with sampling schemes based on simulating a stochastic process in the literature. This was done in a recently published preprint by Thaler et al. \cite{MCMCfailure}. However, they failed to get competitive results while attempting to sample the Bayesian posterior and had to resort to sampling an artificially sharpened distribution. Even then they found that to reliably quantify uncertainty, they had to generate samples from multiple different stochastic processes while we generate samples from a single Markov chain.

\section{The Base Model}
\subsection{The Formal Setting}
The aim of the neural network-based interatomic force model is to map a configuration of atoms $\boldsymbol{x}=\{(\boldsymbol{r}_1,z_1),...,(\boldsymbol{r}_n,z_n)\}$, where $\boldsymbol{r}_i$
 denotes the position of the nucleus of atom $i$ and $z_i$ refers to its nuclear charge, to the forces $\{\boldsymbol{F}_1,...,\boldsymbol{F}_n\}$ acting on the individual nuclei.
These forces are then used to simulate the movement of the nuclei based on Newton's equations of motion via a solver for Ordinary Differential Equations (ODEs) (Figure 1). \\
One important aspect of this form of data is, that it is unstructured, meaning the order in which the atoms are enumerated is arbitrary. A suitable neural network model should therefore be equivariant under a reordering of the data. \\
Another important constraint is, that these forces are conservative and can therefore be derived as the negative gradients of a single potential energy surface $$\boldsymbol{F}_i= -\nabla_{\boldsymbol{r}_i}E((\boldsymbol{r}_1,z_1),...,(\boldsymbol{r}_n,z_n)).$$
%This can be achieved via modern deep learning libraries by having the neural network predict the potential energy and calculating the gradients with those libraries.
The potential energy surface itself is invariant under any distance preserving transformation of the atomic coordinates. This was at first incorporated into neural network models by using only the interatomic distances $r_{ij}$ and nuclear charges as input, which have these invariances themself. Even though it is in principle possible to extract directional information from the set of interatomic distances, in practice incorporating directional information explicitly can improve data efficiency and accuracy quite a lot \cite{Survey_GNNP} and has become a key feature of many state-of-the-art neural network models \cite{GemNet,SpookyNet,PaiNN,NewtonNet,NequIP,UNiTE}.
%\subsection{Atomic Systems as a Graph}
%Because interatomic forces are mostly local, the representation of the input data should also reflect this locality.

 \subsection{The NequIP Model}
\begin{figure*}

\centering
\includegraphics[scale=0.75]{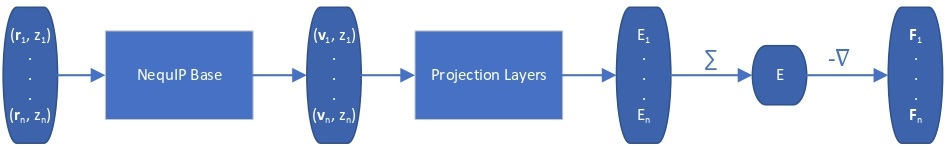}
\caption{The computational graph of the NequIP model for predicting the potential energy $E$ and atomic forces $\boldsymbol{F}_i$ from the atomic numbers $z_i$ and coordinates $\boldsymbol{r}_i$.}
\end{figure*}

One of the most powerful neural network models for interatomic force modeling that currently exists is the NequIP model \cite{NequIP} (Figure 2).
This model takes all the previous considerations into account and consists of a model base and projection layers.
The model base maps an atomic configuration $\{(\boldsymbol{r}_1,z_1),...,(\boldsymbol{r}_n,z_n)\}$ to a set of high dimensional latent feature vectors $\{\boldsymbol{v}_1,...,\boldsymbol{v}_n\}$ that are invariant under distance-preserving transformations. The projection layers then map $\{(\boldsymbol{v}_1,z_1),...,(\boldsymbol{v}_n,z_n)\}$ to (virtual) atomic energies $\{E_1,...,E_n\}$. \\
The potential energy $E$ is then calculated as the sum of the atomic energies. Finally, the forces acting on the nuclei are then calculated as the negative gradients of the potential energy with respect to the nuclear coordinates. The last step can be achieved straightforwardly using the automatic differentiation functionalities of modern deep-learning libraries.

\section{The Bayesian Model}
\subsection{Bayesian Neural Networks}
The goal of classical neural network optimization is to find a single set of parameters, i.e. weights and biases, which models the underlying data distribution, from which the training data was sampled, well.
In the Bayesian approach, in contrast, the parameters of the neural network are modeled probabilistically.
To keep the notation simple, we use $\boldsymbol{\theta}$ to denote a vector containing a complete set of neural network parameters.
For BNNs it is assumed that some prior knowledge exists about what constitutes a good set of parameters, which is expressed in the form of a prior density $p(\boldsymbol{\theta})$.
This prior density then gets refined through the training data $D=\{(\boldsymbol{x}_1,\boldsymbol{y}_1),...,(\boldsymbol{x}_l,\boldsymbol{y}_l)\}$ by invoking Bayes rule for calculating the posterior density:$$p(\boldsymbol{\theta}|D)=\dfrac{p(D|\boldsymbol{\theta})p(\boldsymbol{\theta})}{p(D)}=\dfrac{p(D|\boldsymbol{\theta})p(\boldsymbol{\theta})}{\int p(D|\boldsymbol{\theta})p(\boldsymbol{\theta}) d\boldsymbol{\theta} }.$$
A prediction on a new data point $(\boldsymbol{x},\boldsymbol{y})$ can now be made via:
$$p(\boldsymbol{y}|\boldsymbol{x},D)=\int p(\boldsymbol{y}|\boldsymbol{x},\boldsymbol{\theta})p(\boldsymbol{\theta}|D)d\theta$$
$$=\mathbb{E}_{p(\boldsymbol{\theta}|D)}\left[p(\boldsymbol{y}|\boldsymbol{x},\boldsymbol{\theta})\right].$$
For BNNs this integral is almost always analytically intractable. However, by generating samples $\boldsymbol{\theta}_1,...,\boldsymbol{\theta}_k$ from $p(\boldsymbol{\theta}|D)$ it can be estimated through the law of large numbers as:
$$p(\boldsymbol{y}|\boldsymbol{x},D) \approx \dfrac{1}{k}\sum_{i=1}^k p(\boldsymbol{y}|\boldsymbol{x},\boldsymbol{\theta}_i).$$

 \subsection{The Stochastic Model}
In order to achieve state-of-the-art accuracy combined with a good measure of predictive uncertainty in modeling interatomic forces, a new stochastic neural network model is required, which we will introduce in this section.\\
To build a stochastic model of the data, we make the following assumptions on the conditional independence of the data:
$$p(D)=p(\boldsymbol{y}_1,...,\boldsymbol{y}_l|\boldsymbol{x}_1,...,\boldsymbol{x}_l)p(\boldsymbol{x}_1,...,\boldsymbol{x}_l)$$
$$=p(\boldsymbol{x}_1,...,\boldsymbol{x}_l)\Pi_{i=1}^l p(\boldsymbol{y}_i|\boldsymbol{x}_i).$$
$p(\boldsymbol{y}_i|\boldsymbol{x}_i)$ will be inferred by a neural network as $p(\boldsymbol{y}_i|\boldsymbol{x}_i,\boldsymbol{\theta})$ while $p(\boldsymbol{x}_1,...,\boldsymbol{x}_l)$ depends on the data generation process.

More specifically we model the conditional densities $$p(\boldsymbol{y}|\boldsymbol{x},\boldsymbol{\theta})=p(E,\boldsymbol{F}_1,...,\boldsymbol{F}_n|(\boldsymbol{r}_1,z_1),...,(\boldsymbol{r}_n,z_n),\boldsymbol{\theta})$$ as 
$$\boldsymbol{y|\boldsymbol{x},\theta}\sim N(\mu_E(\boldsymbol{\theta},\boldsymbol{x}),\sigma_E^2(\boldsymbol{\theta},\boldsymbol{x}))$$
$$\;\;\,\;\;\;\;\; \;\;\,\;\;\;\;\;\times \Pi_{i=1}^nN(\boldsymbol{\mu_{\boldsymbol{F}}}_i(\boldsymbol{\theta},\boldsymbol{x}),\sigma_{\boldsymbol{F}_i}^2(\boldsymbol{\theta},\boldsymbol{x})I),$$
where $\boldsymbol{y|\boldsymbol{x},\theta}$ denotes $\boldsymbol{y}$ conditioned on $\boldsymbol{x}$ and $\boldsymbol{\theta}$, $\boldsymbol{x}=\{(\boldsymbol{r}_1,z_1),...,(\boldsymbol{r}_n,z_n)\}$, the means $\mu_E(\boldsymbol{\theta},\boldsymbol{x})$ and $\boldsymbol{\mu_{\boldsymbol{F}}}_i(\boldsymbol{\theta},\boldsymbol{x})$ are the regular point predictions of the NequIP model, the standard deviations $\sigma_E(\boldsymbol{\theta},\boldsymbol{x})$ and $\sigma_{\boldsymbol{F}_i}(\boldsymbol{\theta},\boldsymbol{x})$ are predicted by two separate Multi-Layer Perceptrons (MLPs) from the outputs of the NequIP base $\boldsymbol{v}_1,...,\boldsymbol{v}_n$ and $I$ denotes the identity matrix (Figure 3).
Note that a single invariant standard deviation is calculated for all three force components to ensure equivariance of the predicted density under any distance-preserving transformation of the atomic coordinates.\\
We include the potential energy $E$ here as an additional dependent variable because it can be useful in applications such as the determination of ground state configurations and is calculated as a byproduct of force calculations in DFT anyway.
Lastly, we use the simple Gaussian mean field prior $\boldsymbol{\theta}\sim N(\boldsymbol{0},I)$.
\begin{figure*}

\centering
\includegraphics[scale=0.75]{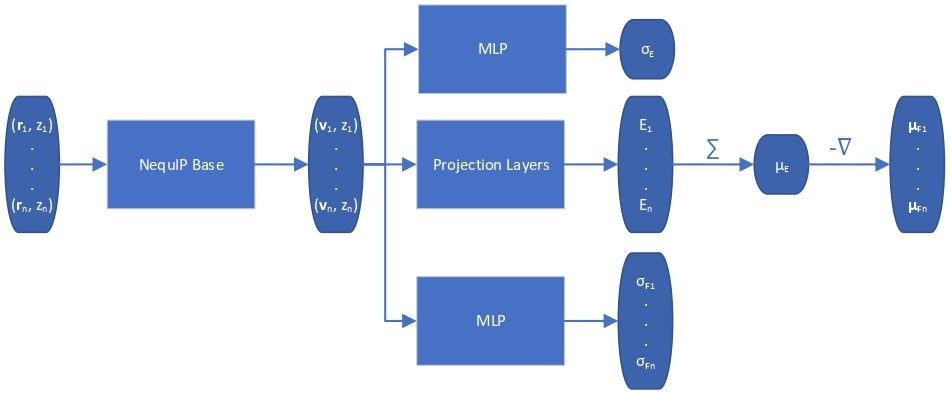}
\caption{The computational graph of the stochastic model for predicting the means $\mu_E,\boldsymbol{\mu}_{\boldsymbol{F}_i}$  and standard deviations $\sigma_{E},\sigma_{\boldsymbol{F}_i}$ of  the potential energy $E$ and atomic forces $\boldsymbol{F}_i$ from the atomic numbers $z_i$ and coordinates $\boldsymbol{r}_i$.}
\end{figure*}

\subsection{Sampling from the Posterior Distribution}

Generating high-quality sample weights from the posterior distribution is usually done via the simulation of a Markov chain which converges in distribution to the posterior.
While several Markov chains methods have been constructed for neural network applications, we found them unsuitable for this use case.
The main difficulty we encountered was, that the scale of the gradients with respect to the different sets of parameters corresponding to the different layers of the neural network vary by several orders of magnitude. This makes the use of a single step size for all parameters impossible, as it would cause the sets of parameters with smaller gradients to be frozen during the optimization. While modern neural network optimizers circumvent this problem through an adaptive step size for each parameter \cite{Optimizers},
this is only possible to a very limited degree for Markov chains \cite{Complete_Recipe} without changing the stationary distribution to which they converge. Furthermore, this typically requires the calculation of higher-order derivatives \cite{Complete_Recipe} which is computationally expensive.
To deal with these challenges, we develop here a new approach to sampling the posterior distribution based on the Stochastic Gradient Hamiltonian Monte Carlo (SGHMC) algorithm introduced by Chen et al. \cite{SGHMC}.
In its basic form, the algorithm is given by the Markov chain\\
$$\Delta \boldsymbol{w}_t= -\left(\nabla_{\boldsymbol{\theta}_t} u(\boldsymbol{\theta}_t) +\boldsymbol{M}^{-1}\boldsymbol{C} \boldsymbol{w}_{t-1} \right)\Delta t+ \sqrt{2 \boldsymbol{C} \Delta t}\boldsymbol{\mathcal{N}}_t(\boldsymbol{0},I),$$
$$\Delta \boldsymbol{\theta}_t =\boldsymbol{M}^{-1}\boldsymbol{w}_t \Delta t .$$
where 
$$u(\boldsymbol{\theta}):= -\ln{p(\boldsymbol{\theta})}-\sum_{i=1}^l\ln{ p(\boldsymbol{y}_i|\boldsymbol{x}_i,\boldsymbol{\theta})},$$
$\boldsymbol{M}^{-1}$ and $\boldsymbol{C}$ are vectors containing strictly positive values, $\Delta t<<1$ is the step size and $\boldsymbol{\mathcal{N}}_t(\boldsymbol{0},I)$ denotes a random variable with a multivariate standard normal distribution. To keep the notation simple, we use the convention here, that all operations on vectors (multiplication, inversion, etc.) are to be taken element-wise. The auxiliary variable $\boldsymbol{w}_t$ has the same dimension as $\boldsymbol{\theta}_t$. \\

A physical intuition of SGHMC can be gained by noticing that these equations also describe the diffusion of particles with coordinates $\boldsymbol{\theta}$, potential energy $u(\boldsymbol{\theta})$, momenta $ \boldsymbol{w}_t$ and masses $\boldsymbol{M}$ under friction. In this analogy, the vector $\boldsymbol{C}$ would represent the friction coefficients.
From a machine learning perspective, $\boldsymbol{w}_t$ represents a momentum term similar to the ones used in many modern neural network optimizers \cite{Optimizers}.
In fact, Chen et al. used some substitutions which lead to a Bayesian analog to stochastic gradient descent with momentum \cite{SGHMC}, but we will make somewhat different substitutions that lead to a natural way to include adaptive step sizes:
$\alpha=\Delta t \boldsymbol{M}^{-1}\boldsymbol{C}$, $\gamma=(\Delta t)^2 |D|\alpha^{-1}$ and $\boldsymbol{v}_t=\gamma^{-1} \Delta t \boldsymbol{w}_t$ which yields:
$$\Delta \boldsymbol{v}_t=-\alpha \frac{1}{|D|} \nabla_{\boldsymbol{\theta}_t} u(\boldsymbol{\theta}_t)-\alpha \boldsymbol{v}_{t-1} +\alpha \sqrt{\frac{2 \boldsymbol{M}}{|D|\gamma}} \boldsymbol{\mathcal{N}}_t\left(0,I\right),$$
$$\Delta \boldsymbol{\theta}_t= \gamma \boldsymbol{M}^{-1} \boldsymbol{v}_t $$
where $|D|$ is the size of the dataset $D$.\\
This is up to the noise term $\boldsymbol{\mathcal{N}}_t\left(0,I\right)$ equivalent to the updates of the Adam optimizer \cite{Adam}. 

Unfortunately, the mass term $\boldsymbol{M}$ used in the standard Adam optimizer can vary a lot even during later stages of the training and this variation depends on the value of $\boldsymbol{\theta}_t$ at the previous time steps. As a consequence, the resulting process $\left(\boldsymbol{\theta}_t,\boldsymbol{v}_t\right)$ would not even be a Markov chain, and the Bayesian posterior would most likely no longer be the distribution $\boldsymbol{\theta}_t$ converges to \cite{Complete_Recipe}.
For many neural network architectures, timely convergence to the posterior can be achieved by simply setting $\boldsymbol{M}=I$. However, the vastly varying gradient scales of the different parameter groups in the model make this approach not feasible.
As a solution, we introduce now a new adaptive step size method for the SGHMC algorithm which still converges to the posterior distribution without requiring the computation of higher-order derivatives.
 In order to achieve this, we set $\boldsymbol{M}$ as the denominator of the AMSGrad algorithm \cite{AMSGrad} during the first phase of the optimization:
$$\boldsymbol{a}_t=(1-\beta) \frac{1}{|D|^2}\left(\nabla_{\boldsymbol{\theta}_t}u(\boldsymbol{\theta}_t)\right) \left(\nabla_{\boldsymbol{\theta}_t}u(\boldsymbol{\theta}_t)\right)+\beta \boldsymbol{a}_{t-1},$$
$$\boldsymbol{D}_t=\max(\boldsymbol{D}_{t-1},\boldsymbol{a}_t),$$
$$\boldsymbol{M}_t=\sqrt{\frac{\boldsymbol{D}_t}{1-\beta^t}}+\epsilon_{stability}.$$
Here, $\max(\cdot,\cdot)$ is the element-wise maximum, $\epsilon_{stability}$ is a stability constant and $\beta$ is a hyperparameter used in computing the running average $\boldsymbol{a}_t$ and is typically set between $0.99$ and $0.9999$. \\
Because this mass term is still time- and path-dependent,  $\boldsymbol{\theta}_t$ can not, in general, be expected to converge exactly to the posterior distribution during this phase.
However, because $\boldsymbol{M}_t$ is not based on a running average of squared gradients, like most adaptive step size methods are, but instead on the maximum of such a running average, it typically changes very little during the later stages of the optimization. As a result, the process will already become fairly close in distribution to the Bayesian posterior during this stage of the optimization.
Furthermore, because $\boldsymbol{M}_t$ already remains almost constant after a while, we can keep it entirely constant after a certain amount of steps without causing instabilities, at which point the process $\boldsymbol{\theta}_t$  becomes a regular SGHMC process which is known to converge to the Bayesian posterior for sufficiently small step sizes \cite{SGHMC}.

\begin{figure*}
\includegraphics[scale=0.55]{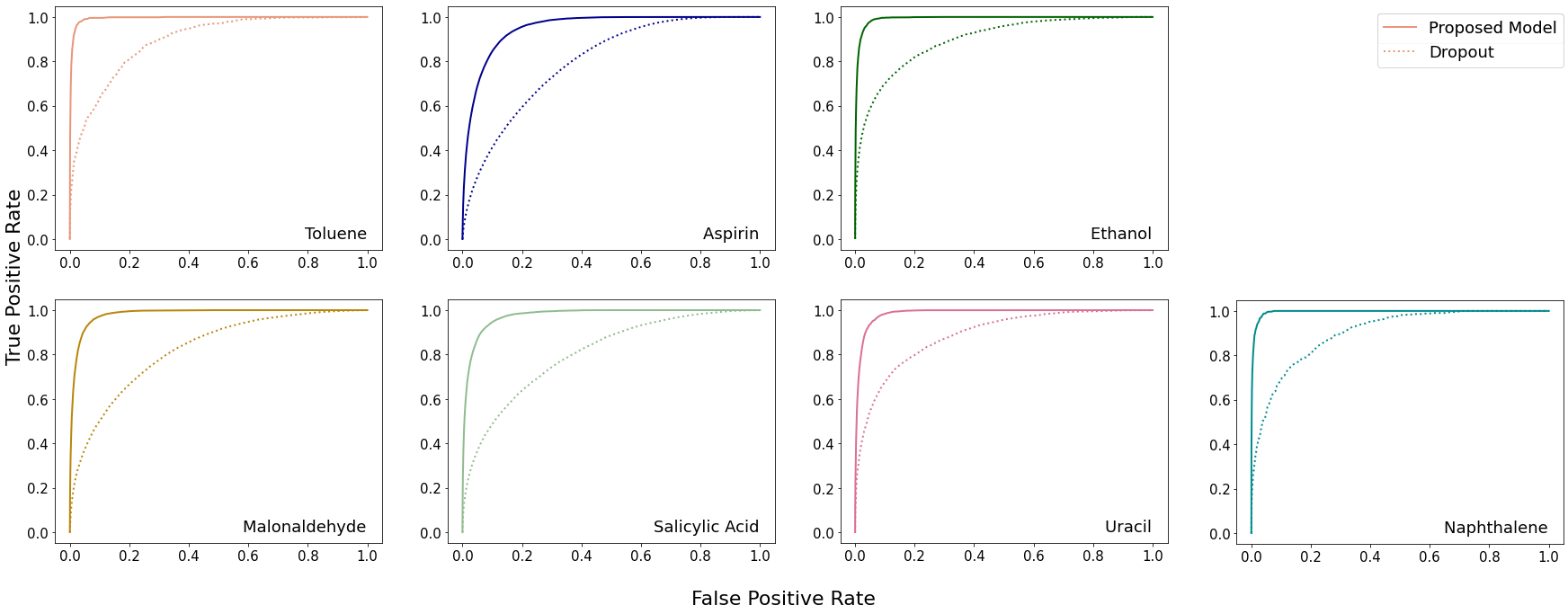}
\caption{Receiver Operating Characteristic curves for uncertainty-based detection of force components with a prediction error of at least 1 kcal/(mol$\cdot$\AA) using k=8 Monte Carlo samples on the RMD17 datasets.}
\end{figure*}

Because deep learning datasets are usually very large, evaluating $ \nabla_{\boldsymbol{\theta}} u(\boldsymbol{\theta})$ exactly is typically very time-consuming.
Practical implementations instead estimate $\nabla_{\boldsymbol{\theta}} u(\boldsymbol{\theta})$ on a smaller, randomly sampled subset $\hat{D} \subset D $ via 
$$\nabla_{\boldsymbol{\theta}} u(\boldsymbol{\theta}) =\boldsymbol{\epsilon}( \boldsymbol{\theta})+ \nabla_{\boldsymbol{\theta}} \hat{u}(\boldsymbol{\theta})$$
$$:=\boldsymbol{\epsilon}( \boldsymbol{\theta})+\nabla_{\boldsymbol{\theta}}\left(-\ln{p(\boldsymbol{\theta})}-\frac{|D|}{|\hat{D}|} \sum_{(\boldsymbol{x},\boldsymbol{y}) \in \hat{D}}\ln{ p(\boldsymbol{y}|\boldsymbol{x},\boldsymbol{\theta})}\right).$$%+\boldsymbol{N}(0,\boldsymbol{\Sigma}(\boldsymbol{\theta}))$$
The resulting algorithm is summarized in \cref{alg:example}.
Here $\boldsymbol{\epsilon}( \boldsymbol{\theta})$ is the error in the estimation of $\nabla_{\boldsymbol{\theta}} u(\boldsymbol{\theta})$ with $\mathbb{E}\left[\boldsymbol{\epsilon}( \boldsymbol{\theta})\right]=\boldsymbol{0}$.\\
\begin{algorithm}[h]
   \caption{The Stochastic Optimization Algorithm}
   \label{alg:example}
\begin{algorithmic}

   \FOR{$t=1$ {\bfseries to} $T$}
   \STATE Sample minibatch $\hat{D}$
   \STATE Evaluate $\nabla_{\boldsymbol{\theta}_t} \hat{u}(\boldsymbol{\theta}_t)$\\
$=\nabla_{\boldsymbol{\theta}_t}\left(-\ln{p(\boldsymbol{\theta}_t)}-\frac{|D|}{|\hat{D}|} \sum_{(\boldsymbol{x},\boldsymbol{y}) \in \hat{D}}\ln{ p(\boldsymbol{y}|\boldsymbol{x},\boldsymbol{\theta}_t)}\right)$
   \IF{$t \leq t_{max}$}

   \STATE $\boldsymbol{a}_t=(1-\beta) \frac{1}{|D|^2}\left(\nabla_{\boldsymbol{\theta}_t}\hat{u}(\boldsymbol{\theta}_t)\right) \left(\nabla_{\boldsymbol{\theta}_t}\hat{u}(\boldsymbol{\theta}_t)\right)+\beta \boldsymbol{a}_{t-1}$
\STATE $\boldsymbol{D}_t=\max(\boldsymbol{D}_{t-1},\boldsymbol{a}_t)$
\STATE $\boldsymbol{M}_t=\sqrt{\frac{\boldsymbol{D}_t}{1-\beta^t}}+\epsilon_{stability}$
\ELSE
\STATE $\boldsymbol{M}_t=\boldsymbol{M}_{t_{max}}$
\ENDIF
\STATE $\Delta \boldsymbol{v}_t=-\frac{\alpha}{|D|} \nabla_{\boldsymbol{\theta}_t} \hat{u}(\boldsymbol{\theta}_t)-\alpha \boldsymbol{v}_{t-1}  +\alpha \sqrt{\frac{2 \boldsymbol{M}_t}{|D|\gamma}} \boldsymbol{\mathcal{N}}_t\left(0,I\right)$
\STATE$\Delta \boldsymbol{\theta}_t= \gamma \boldsymbol{M}_t^{-1} \boldsymbol{v}_t $
\ENDFOR
  % \UNTIL{$noChange$ is $true$}
\end{algorithmic}
\end{algorithm}
As long as $\frac{|\boldsymbol{\epsilon}( \boldsymbol{\theta})|}{|D|} << \sqrt{\frac{2 \boldsymbol{M}_t}{\gamma |D|}}$ it is clear that this additional noise does not have a large effect on the dynamic as the Gaussian noise term will dominate.\\
This can always be achieved by choosing an appropriate batch size and step size $\gamma$. \\
However, the additional noise will become significant for large learning rates and small batch sizes.\\ 
In fact, even if we do not add the Gaussian noise to this process and subsequently just use the resulting AMSGrad optimization, the injected noise at the commonly used large step sizes and small batch sizes will cause the optimization process to become a stochastic process which imitates the proposed algorithm to a degree, only with a noise term $\frac{|\boldsymbol{w}( \boldsymbol{\theta})|}{|D|}$ that is not grounded on any theoretical foundation.
\section{Empirical Evaluation}
\subsection{The Benchmarks}

To assess our model's accuracy and predicted uncertainty we utilize two different datasets.
To analyze our model's performance with varying amounts of Monte Carlo samples and under domain shift, we used a dataset consisting of PEDOT polymers, which are conducting polymers, which due to their chemical stability and tuneable properties \cite{Ben1} have been used for a wide range of applications including sensors \cite{Ben2}, supercapacitors \cite{Ben3}, battery electrodes \cite{Ben4}, bioelectronics, solar cells, electrochromic displays, electrochemical transistors \cite{Ben5}, and spintronics \cite{Ben6}.  The dataset consists of polymers of lengths 8, 12 and 16, where only the shorter chains were used for training (see appendix B for details). The total training set contained only 100 configurations, 50 of length 8 polymers and 50 of length 12 polymers. Equivalently a small validation set of size 30 was constructed.

The second dataset used is the MD17 dataset \cite{MD17} consisting of long molecular dynamics trajectories of several small organic molecules calculated in DFT. Because this dataset had some numerical errors, a revised version RMD17 has recently been computed at higher accuracy \cite{RMD17} which we will utilize to evaluate our model's performance. Because dropout is the most common Bayesian method used for interatomic force modeling, we will compare our model's performance to a dropout-based version of the proposed stochastic model on this dataset (see appendix A.4 for details). \\
For each compound, we used 1000 randomly sampled configurations during training and the rest for testing. Of the 1000 configurations sampled, 30 were reserved as a small validation set, and the remaining ones comprised the actual training set.
The same samples were used as training, validation and test sets for our proposed model and the dropout-based model.

\begin{figure*}
\includegraphics[scale=0.55]{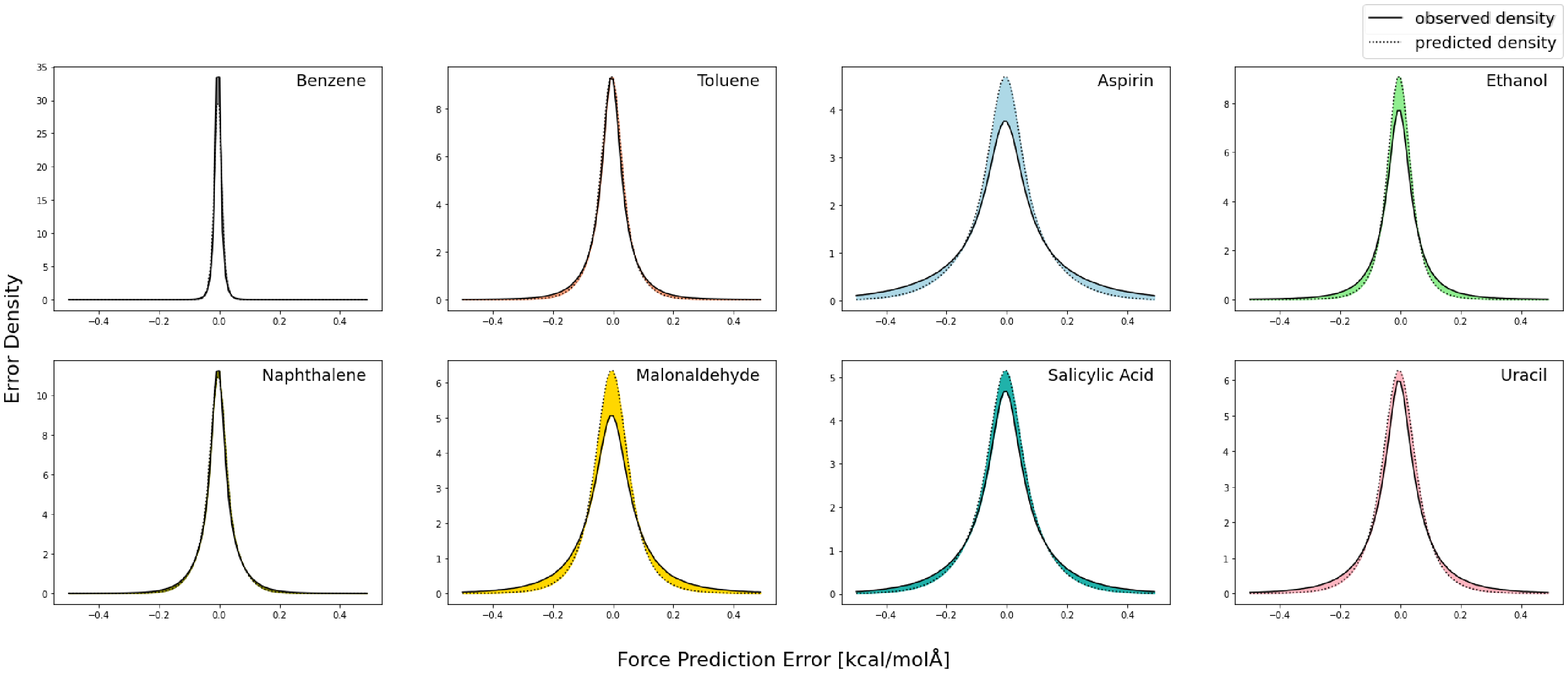}
\caption{Observed and predicted error densities of the force components in kcal/(mol$\cdot$\AA) with k=8 Monte Carlo samples on the RMD17 datasets. }
\end{figure*}

\subsection{Evaluation Metrics}
To measure prediction accuracy, we evaluate both the Mean Average Error in the Forces (MAE-F) and Energies (MAE-E) as well as the Root Mean Square Errors (RMSE-F, RMSE-E).
In the evaluation, we use the expectation value under the estimated posterior density as a point prediction. I.e. $\boldsymbol{F}_{pred}(\boldsymbol{x})= \frac{1}{k}\sum_{i=1}^k \boldsymbol{\mu}_{\boldsymbol{F}}(\boldsymbol{x},\boldsymbol{\theta}_i)$ and $E_{pred}(\boldsymbol{x})= \frac{1}{k}\sum_{i=1}^k \mu_E(\boldsymbol{x},\boldsymbol{\theta}_i)$.
One metric used to evaluate and compare the predicted uncertainties is the Mean Log Likelihood (MLL). Since in the most common applications, the main task of the uncertainty measure is outlier detection, we further evaluate the ROC AUC scores for detecting force components with an error larger than 1 kcal/(mol$\cdot $\AA) on the basis of the variance of the predicted distribution of those force components. Because of the relatively small size of the PEDOT dataset, we only evaluate the outlier detection on the much larger RMD17 datasets.

Lastly, to evaluate if the proposed model is properly calibrated we utilize the Expected Calibration Error (ECE) \cite{Binning} as well as a visual inspection of the predicted and observed error densities of the force components (Figures 4 and 5) (see appendix A.5 for details). To calculate an ECE, the predictions $y_i$ are divided into $m$ bins of equal width. The ECE is then calculated as
$$ECE=\sum_{i=1}^m f_i\left|f_i-e_i\right|$$
where $f_i$ is the fraction of observed samples that fall into bin $i$ and $e_i$ is the predicted probability of a sample falling into bin $i$. Because the ECE is highly dependent on the width of the bins, we divide the ECE by the bin width $\delta$ to obtain a Normalized Expectation Calibration Error (NECE):
$$NECE=\sum_{i=1}^m \frac{f_i}{\delta}\left|f_i-e_i\right|$$
which is an estimate of 
$$\mathbb{E}_{\rho(y)}\left[|\rho(y)-\rho_{predicted}(y)|\right].$$

\subsection{Results}
\paragraph{Results on the PEDOT Datasets\\}

As can be seen in Table 2, the model achieves high accuracy in the force prediction on all three test sets with a decreasing accuracy for increasing chain lengths. A slight improvement in the accuracy is observed when going from one to eight Monte Carlo samples. The model has a much larger error in the auxiliary task of energy predictions when compared to the force predictions. This is not surprising though, as the PEDOT molecules are fairly large and have correspondingly large energy scales. Additionally, the PEDOT training set contains much fewer energy labels than force labels.
Further, it can be seen from Table 3, that a single Monte Carlo sample does not yield good uncertainty estimates for the forces which vastly improves when using eight Monte Carlo samples. A similar trend is not evident for the uncertainty quantification of the energy predictions.

Even though no polymer chains of length 16 were included in the training set, we find that the model still achieves high accuracy and a good quantification of the uncertainty in the force predictions (Table 3) in the case of k=8 Monte Carlo samples and again much poorer uncertainty quantification for a single Monte Carlo sample. A complete histogram of predicted and actual force components can be found in appendix C. We observe some overconfidence even in the k=8 case (Figure 6). 

\begin{figure}[h]
\centering
\includegraphics[scale=0.5]{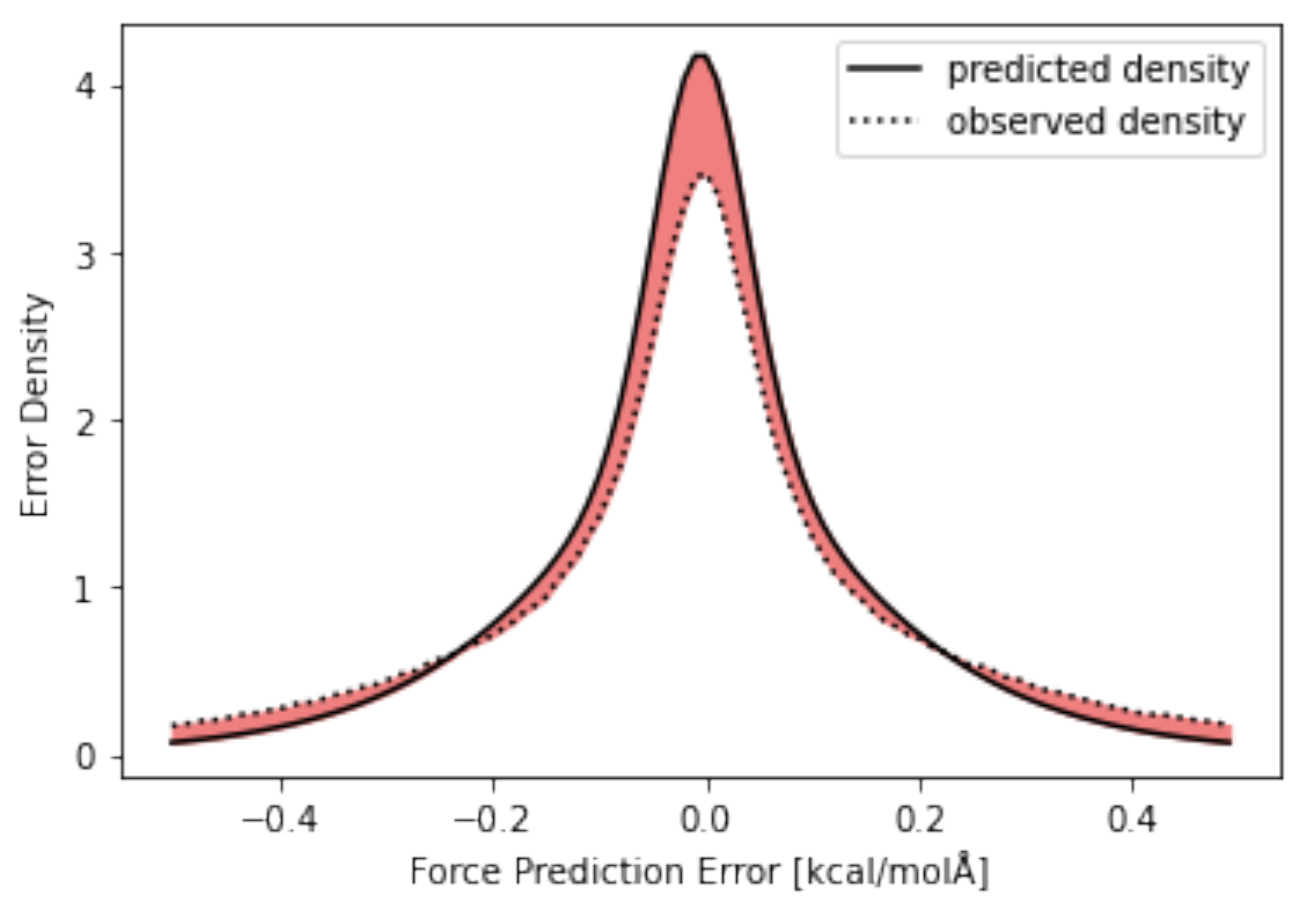}
\caption{Observed and predicted error densities of the force components in kcal/(mol$\cdot$\AA) with k=8 Monte Carlo samples on the length 16 PEDOT dataset. }
\end{figure}

\paragraph{Influence of the Number of Monte Carlo Samples\\}
\begin{table*}[h]
%\begin{centering}
\centering
\caption{Results on the PEDOT datasets. Force in kcal/(mol$\cdot$\AA), energy in kcal/mol.}
\includegraphics[scale=0.8]{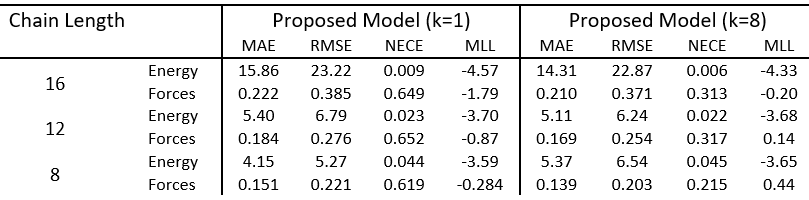}

%\end{centering}
\end{table*}
As can be seen in Table 3, both the calibration as well as the MLL significantly improve with increasing sample size for the forces. Because these samples are generated from a single Markov chain, this demonstrates an efficient traversal of the parameter space by the proposed sampling algorithm. Further, it appears that the improvements in MLLs diminish exponentially with increasing sample size. No similar trend can be found for the auxiliary task of energy predictions. Overall we find $k=8$ Monte Carlo samples to be a good tradeoff between quality of uncertainty quantification and computational complexity.
\begin{table}[h]
\caption{Influence of the MC sample size on the length 16 PEDOT dataset. Force in kcal/(mol$\cdot$\AA), energy in kcal/mol.}
\centering
\includegraphics[scale=0.85]{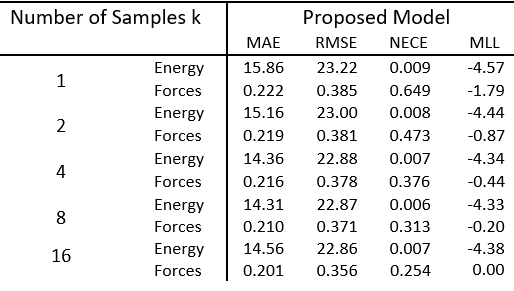}

\end{table}

\paragraph{Results on the RMD17 Datasets\\}
Using eight Monte Carlo samples for both models, the proposed model has comparable accuracy to the dropout model on many of the easier RMD17 datasets but significantly outperforms it on the aspirin and malonaldehyde datasets (Table 4). The achieved accuracies are very consistent with the original NequIP model on these datasets \cite{NequIP} (See appendix D for the results of a single Monte Carlo sample and the NequIP model on this dataset). Further, it consistently outperforms the dropout model in terms of mean log-likelihoods for both the forces and the energies. The only exception to this being the mean log-likelihoods of the uracil energies where both models achieve the same result.

\begin{table}
%\begin{centering}

 \caption{Mean results on the RMD17 dataset using $k=8$ Monte Carlo samples. Force in kcal/(mol$\cdot$\AA), energy in kcal/mol.}
\centering
 \includegraphics[scale=0.75]{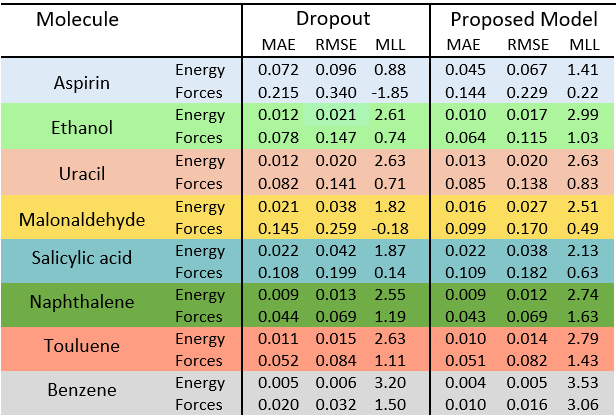}
%\end{centering}
\end{table}

A very large difference in the performance of the models is found in the outlier detection task where the proposed model consistently achieves ROC AUC scores much closer to the optimal score of 1 (Table 5). The benzene dataset was not included in this comparison because there were no instances of force prediction errors of the necessary scale for the proposed model. A complete plot of the receiver operating characteristic curves is given in Figure 4. As can be seen there, our proposed model is much more reliable at detecting outliers.
\begin{table}
%\begin{centering}
 \caption{ROC AUC scores for uncertainty-based detection of force components with a prediction error of at least 1 kcal/(mol$\cdot$\AA) using k=8 Monte Carlo samples on the RMD17 datasets.}
\centering
 \includegraphics[scale=0.75]{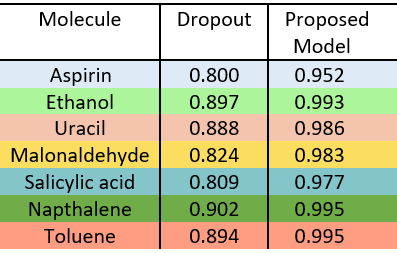}
%\end{centering}
\end{table}

Lastly, as is evident from Table 6 and Figure 5 the resulting model is not accurately calibrated in the case of 8 Monte Carlo samples and has a tendency for overconfidence.
\begin{table}
%\begin{centering}
 \caption{Calibration errors on the RMD17 dataset using k=8 Monte Carlo samples}
\centering
 \includegraphics[scale=0.75]{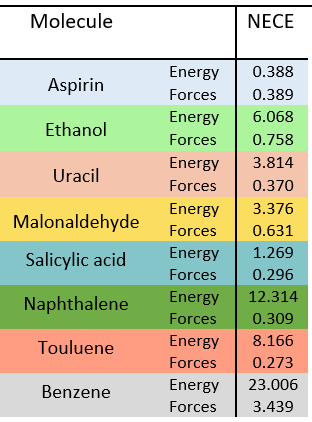}
%\end{centering}
\end{table}

\section{Conclusion and Outlook}
As the results demonstrate, the proposed Bayesian neural network model achieves state-of-the-art accuracy while also achieving good uncertainty quantification. The proposed sampling algorithm appears to converge to the posterior distribution in a reasonable amount of time and maintains a fairly swift traversal through the parameter space afterward. Most importantly, the resulting model can reliably detect outliers while sampling parameters from the same Markov chain. This opens up new possibilities for Bayesian active learning procedures for learning interatomic forces. Further, it could enable the principled incorporation of existing physics-guided approximate interatomic force fields via the Bayesian prior distribution. This might potentially further improve both the data efficiency and generalizability of the model.\\
While the proposed model already achieves good results, some improvements could still be made.\\
While convergence to the posterior distribution becomes feasible with the proposed algorithm, in practice we find that it still takes about twice as many GPU hours to reach convergence when compared to the non-Bayesian optimization procedure described in \cite{NequIP}. However, while it was useful for demonstrative purposes of the convergence properties of the proposed sampler, faster convergence can most likely be achieved by simply reducing the injected gradient noise and the batch size during the initial phases of the optimization.
Nevertheless, even without these measures, the generation of interatomic forces for the training set at {\it ab initio} accuracy will likely be the computational bottleneck in most applications.\\
 Further, the stochastic model could potentially still be improved by adequate incorporation of covariances between atoms and also between force components. However, both of these covariances are challenging to include. For covariance between force components, the main challenge is to maintain rotation equivariance of the predicted density. For the incorporation of interatomic covariances, a dense covariance matrix will very quickly become impractical for larger molecules, as the evaluation of the log-likelihoods becomes a computational bottleneck. While a sparse covariance matrix might be adequate due to the mostly local nature of interatomic forces, a way to parameterize sparse covariance matrices would be needed, which is not trivial and which we could not find in the existing literature.\\
Lastly, we found that the predicted uncertainties are not always properly calibrated and have a tendency for overconfidence. This might be a result of sampling from the same Markov chain, where the models can never be completely independently sampled from each other. This can lead to a reduced predictive variance between the sampled models and hence increased confidence. Here it might be beneficial to utilize two or more parallel chains to reduce the sample interdependency and explore the posterior distribution more efficiently or to recalibrate the uncertainties on a validation set.
\section{Acknowledgement}
This paper is funded by dtec.bw – Digitalization and Technology Research Center of the Bundeswehr which we gratefully acknowledge [project CoupleIT!]
\bibliography{bib}

\begin{thebibliography}{10}

\bibitem{Atkins}
P.W. Atkins and R.S. Friedman.
\newblock {\em Molecular Quantum Mechanics}.
\newblock OUP Oxford, 2011.

\bibitem{MD_NNP}
M.~Gastegger and P.~Marquetand.
\newblock Molecular dynamics with neural network potentials.
\newblock In {\em Machine Learning Meets Quantum Physics}, pages 233--252.
  Springer, 2020.

\bibitem{Survey_NNP}
Emir Kocer, Tsz~Wai Ko, and J\"{o}rg Behler.
\newblock Neural network potentials: A concise overview of methods.
\newblock {\em Annual Review of Physical Chemistry}, 73(1):163--186, 2022.
\newblock PMID: 34982580.

\bibitem{GemNet}
Johannes Klicpera, Florian Becker, and Stephan G{\"u}nnemann.
\newblock Gemnet: Universal directional graph neural networks for molecules.
\newblock In A.~Beygelzimer, Y.~Dauphin, P.~Liang, and J.~Wortman Vaughan,
  editors, {\em Advances in Neural Information Processing Systems}, 2021.

\bibitem{SpookyNet}
Oliver~T. Unke, Stefan Chmiela, Michael Gastegger, Kristof~T. Schütt,
  Huziel~E. Sauceda, and Klaus-Robert Müller.
\newblock Spookynet: Learning force fields with electronic degrees of freedom
  and nonlocal effects.
\newblock {\em Nature Communications}, 12, 2021.

\bibitem{PaiNN}
Kristof~T. Schütt, Oliver~T. Unke, and Michael Gastegger.
\newblock Equivariant message passing for the prediction of tensorial
  properties and molecular spectra.
\newblock {\em CoRR}, abs/2102.03150, 2021.

\bibitem{NequIP}
Simon Batzner, Albert Musaelian, Lixin Sun, Mario Geiger, Jonathan~P. Mailoa,
  Mordechai Kornbluth, Nicola Molinari, Tess~E. Smidt, and Boris Kozinsky.
\newblock E(3)-equivariant graph neural networks for data-efficient and
  accurate interatomic potentials.
\newblock {\em Nature Communications}, 13, 2022.

\bibitem{NewtonNet}
M.~Haghighatlari, J.~Li, X.~Guan, O.~Zhang, A.~Das, C.~J. Stein,
  F.~Heidar-Zadeh, M.~Liu, M.~Head-Gordon, L.~Bertels, H.~Hao, I.~Leven, and
  T.~Head-Gordon.
\newblock Newtonnet: a {N}ewtonian message passing network for deep learning of
  interatomic potentials and forces.
\newblock {\em Digital discovery}, 1(3):333–343, 2022.

\bibitem{UNiTE}
Zhuoran Qiao, Anders~S. Christensen, Matthew Welborn, Frederick~R. Manby, Anima
  Anandkumar, and Thomas~F. Miller.
\newblock Informing geometric deep learning with electronic interactions to
  accelerate quantum chemistry.
\newblock {\em Proceedings of the National Academy of Sciences},
  119(31):e2205221119, 2022.

\bibitem{Survey_GNNP}
P.~AU~Reiser, M.~Neubert, A.~Eberhard, L.~Torresi, C.~Zhou, C.~Shao, H.~Metni,
  C.~van Hoesel, H.~Schopmans, T.~Sommer, and P.~Friederich.
\newblock Graph neural networks for materials science and chemistry.
\newblock {\em Communications Materials}, 3(93), 2022.

\bibitem{AL3}
Evgeny~V. Podryabinkin and Alexander~V. Shapeev.
\newblock Active learning of linearly parametrized interatomic potentials.
\newblock {\em Computational Materials Science}, 140:171--180, 2017.

\bibitem{AL2}
Evgeny~V. Podryabinkin, Evgeny~V. Tikhonov, Alexander~V. Shapeev, and Artem~R.
  Oganov.
\newblock Accelerating crystal structure prediction by machine-learning
  interatomic potentials with active learning.
\newblock {\em Phys. Rev. B}, 99:064114, 2019.

\bibitem{AL4}
Konstantin Gubaev, Evgeny~V. Podryabinkin, Gus~L.W. Hart, and Alexander~V.
  Shapeev.
\newblock Accelerating high-throughput searches for new alloys with active
  learning of interatomic potentials.
\newblock {\em Computational Materials Science}, 156:148--156, 2019.

\bibitem{AL1}
Ryosuke Jinnouchi, Kazutoshi Miwa, Ferenc Karsai, Georg Kresse, and Ryoji
  Asahi.
\newblock On-the-fly active learning of interatomic potentials for large-scale
  atomistic simulations.
\newblock {\em The Journal of Physical Chemistry Letters}, 11:6946–6955,
  2020.

\bibitem{MCMC_BNNP}
A.~Shapeev, K.~Gubaev, E.~Tsymbalov, and E.~Podryabinkin.
\newblock Active learning and uncertainty estimation.
\newblock In {\em Machine Learning Meets Quantum Physics}, pages 309--329.
  Springer, 2020.

\bibitem{BNNP_DO1}
Mingjian {Wen} and Ellad~B. {Tadmor}.
\newblock {Uncertainty quantification in molecular simulations with dropout
  neural network potentials}.
\newblock {\em npj Computational Mathematics}, 6:124, 2020.

\bibitem{BNNP_DO2}
Sung-Ho Lee, Valerio Olevano, and Benoit Sklénard.
\newblock A generalizable, uncertainty-aware neural network potential for
  gesbte with monte carlo dropout.
\newblock {\em Solid-State Electronics}, page 108508, 2022.

\bibitem{MCMCfailure}
G.~Thaler, S.and~Doehner and J.~Zavadlav.
\newblock Scalable bayesian uncertainty quantification for neural network
  potentials: Promise and pitfalls.
\newblock {\em arxiv}, 2022.

\bibitem{BNNP_DO}
Yarin Gal and Zoubin Ghahramani.
\newblock Dropout as a {B}ayesian approximation: Representing model uncertainty
  in deep learning.
\newblock In {\em Proceedings of the 33rd International Conference on
  International Conference on Machine Learning - Volume 48}, ICML'16, page
  1050–1059. JMLR.org, 2016.

\bibitem{BNN_quality1}
J.~Yao, W.~Pan, S.~Ghosh, and F.~Doshi-Velez.
\newblock Quality of uncertainty quantification for {B}ayesian neural network
  inference.
\newblock In {\em proceedings at the International Conference on Machine
  Learning: Workshop on Uncertainty \& Robustness in Deep Learning (ICML)},
  2019.

\bibitem{Optimizers}
Sebastian Ruder.
\newblock An overview of gradient descent optimization algorithms.
\newblock {\em ArXiv}, abs/1609.04747, 2016.

\bibitem{Complete_Recipe}
Yi-An Ma, Tianqi Chen, and Emily Fox.
\newblock A complete recipe for stochastic gradient mcmc.
\newblock In C.~Cortes, N.~Lawrence, D.~Lee, M.~Sugiyama, and R.~Garnett,
  editors, {\em Advances in Neural Information Processing Systems}, volume~28.
  Curran Associates, Inc., 2015.

\bibitem{SGHMC}
Tianqi Chen, Emily~B. Fox, and Carlos Guestrin.
\newblock Stochastic gradient {H}amiltonian monte carlo.
\newblock In {\em ICML}, pages 1683--1691, 2014.

\bibitem{Adam}
Diederik~P. Kingma and Jimmy Ba.
\newblock Adam: {A} method for stochastic optimization.
\newblock In Yoshua Bengio and Yann LeCun, editors, {\em 3rd International
  Conference on Learning Representations, {ICLR} 2015, San Diego, CA, USA, May
  7-9, 2015, Conference Track Proceedings}, 2015.

\bibitem{AMSGrad}
Sashank~J. Reddi, Satyen Kale, and Sanjiv Kumar.
\newblock On the convergence of adam and beyond.
\newblock In {\em International Conference on Learning Representations}, 2018.

\bibitem{Ben1}
Magatte~N. Gueye, Alexandre Carella, Nicolas Massonnet, Etienne Yvenou, Sophie
  Brenet, Jérôme Faure-Vincent, Stéphanie Pouget, François Rieutord, Hanako
  Okuno, Anass Benayad, Renaud Demadrille, and Jean-Pierre Simonato.
\newblock Structure and dopant engineering in pedot thin films: Practical tools
  for a dramatic conductivity enhancement.
\newblock {\em Chemistry of Materials}, 28(10):3462--3468, 2016.

\bibitem{Ben2}
Thanh-Hai Le, Yukyung Kim, and Hyeonseok Yoon.
\newblock Electrical and electrochemical properties of conducting polymers.
\newblock {\em Polymers}, 9(4), 2017.

\bibitem{Ben3}
Magatte~N. Gueye, Alexandre Carella, Jérôme Faure-Vincent, Renaud Demadrille,
  and Jean-Pierre Simonato.
\newblock Progress in understanding structure and transport properties of
  pedot-based materials: A critical review.
\newblock {\em Progress in Materials Science}, 108:100616, 2020.

\bibitem{Ben4}
Nicholas~S. Hudak.
\newblock Chloroaluminate-doped conducting polymers as positive electrodes in
  rechargeable aluminum batteries.
\newblock {\em The Journal of Physical Chemistry C}, 118(10):5203--5215, 2014.

\bibitem{Ben5}
Igor Zozoulenko, Amritpal Singh, Sandeep~Kumar Singh, Viktor Gueskine, Xavier
  Crispin, and Magnus Berggren.
\newblock Polarons, bipolarons, and absorption spectroscopy of pedot.
\newblock {\em ACS Applied Polymer Materials}, 1(1):83--94, 2019.

\bibitem{Ben6}
Kazuya Ando, Shun Watanabe, Sebastian Mooser, Eiji Saitoh, and Henning
  Sirringhaus.
\newblock Solution-processed organic spin–charge converter.
\newblock {\em Nature Materials}, 12:622–627, 2013.

\bibitem{MD17}
Stefan Chmiela, Alexandre Tkatchenko, Huziel~E. Sauceda, Igor Poltavsky,
  Kristof~T. Schütt, and Klaus-Robert Müller.
\newblock Machine learning of accurate energy-conserving molecular force
  fields.
\newblock {\em Science Advances}, 3(5):e1603015, 2017.

\bibitem{RMD17}
Anders~S. Christensen and O.~Anatole von Lilienfeld.
\newblock On the role of gradients for machine learning of molecular energies
  and forces.
\newblock {\em Machine Learning: Science and Technology}, 1, 2020.

\bibitem{Binning}
Mahdi~Pakdaman Naeini, Gregory~F. Cooper, and Milos Hauskrecht.
\newblock Obtaining well calibrated probabilities using {B}ayesian binning.
\newblock In {\em Proceedings of the Twenty-Ninth AAAI Conference on Artificial
  Intelligence}, AAAI'15, page 2901–2907. AAAI Press, 2015.

\bibitem{e3nn}
Mario Geiger, Tess Smidt, Alby M., Benjamin~Kurt Miller, Wouter Boomsma,
  Bradley Dice, Kostiantyn Lapchevskyi, Maurice Weiler, Michał Tyszkiewicz,
  Martin Uhrin, Simon Batzner, Dylan Madisetti, Jes Frellsen, Nuri Jung, Sophia
  Sanborn, jkh, Mingjian Wen, Josh Rackers, Marcel Rød, and Michael Bailey.
\newblock e3nn/e3nn: 2022-12-12, 2022.

\bibitem{CMCMC}
Ruqi Zhang, Chunyuan Li, Jianyi Zhang, Changyou Chen, and Andrew~Gordon Wilson.
\newblock Cyclical stochastic gradient mcmc for {B}ayesian deep learning.
\newblock In {\em International Conference on Learning Representations}, 2020.

\bibitem{Ben7}
E.~Aprà, E.~J. Bylaska, W.~A. de~Jong, N.~Govind, K.~Kowalski, T.~P.
  Straatsma, M.~Valiev, H.~J.~J. van Dam, Y.~Alexeev, J.~Anchell, V.~Anisimov,
  F.~W. Aquino, R.~Atta-Fynn, J.~Autschbach, N.~P. Bauman, J.~C. Becca, D.~E.
  Bernholdt, K.~Bhaskaran-Nair, S.~Bogatko, P.~Borowski, J.~Boschen, J.~Brabec,
  A.~Bruner, E.~Cauët, Y.~Chen, G.~N. Chuev, C.~J. Cramer, J.~Daily, M.~J.~O.
  Deegan, T.~H. Dunning, M.~Dupuis, K.~G. Dyall, G.~I. Fann, S.~A. Fischer,
  A.~Fonari, H.~Früchtl, L.~Gagliardi, J.~Garza, N.~Gawande, S.~Ghosh,
  K.~Glaesemann, A.~W. Götz, J.~Hammond, V.~Helms, E.~D. Hermes, K.~Hirao,
  S.~Hirata, M.~Jacquelin, L.~Jensen, B.~G. Johnson, H.~Jónsson, R.~A.
  Kendall, M.~Klemm, R.~Kobayashi, V.~Konkov, S.~Krishnamoorthy, M.~Krishnan,
  Z.~Lin, R.~D. Lins, R.~J. Littlefield, A.~J. Logsdail, K.~Lopata, W.~Ma,
  A.~V. Marenich, J.~Martin~del Campo, D.~Mejia-Rodriguez, J.~E. Moore, J.~M.
  Mullin, T.~Nakajima, D.~R. Nascimento, J.~A. Nichols, P.~J. Nichols,
  J.~Nieplocha, A.~Otero-de-la Roza, B.~Palmer, A.~Panyala, T.~Pirojsirikul,
  B.~Peng, R.~Peverati, J.~Pittner, L.~Pollack, R.~M. Richard, P.~Sadayappan,
  G.~C. Schatz, W.~A. Shelton, D.~W. Silverstein, D.~M.~A. Smith, T.~A. Soares,
  D.~Song, M.~Swart, H.~L. Taylor, G.~S. Thomas, V.~Tipparaju, D.~G. Truhlar,
  K.~Tsemekhman, T.~Van~Voorhis, Á. Vázquez-Mayagoitia, P.~Verma, O.~Villa,
  A.~Vishnu, K.~D. Vogiatzis, D.~Wang, J.~H. Weare, M.~J. Williamson, T.~L.
  Windus, K.~Woliński, A.~T. Wong, Q.~Wu, C.~Yang, Q.~Yu, M.~Zacharias,
  Z.~Zhang, Y.~Zhao, and R.~J. Harrison.
\newblock Nwchem: Past, present, and future.
\newblock {\em The Journal of Chemical Physics}, 152(18):184102, 2020.

\bibitem{Ben9}
Axel~D. Becke.
\newblock Density‐functional thermochemistry. iii. the role of exact
  exchange.
\newblock {\em The Journal of Chemical Physics}, 98(7):5648--5652, 1993.

\bibitem{Ben8}
John~P. Perdew, Matthias Ernzerhof, and Kieron Burke.
\newblock Rationale for mixing exact exchange with density functional
  approximations.
\newblock {\em The Journal of Chemical Physics}, 105(22):9982--9985, 1996.

\bibitem{Ben10}
Vitaly~A. {Rassolov}, John~A. {Pople}, Mark~A. {Ratner}, and Theresa~L.
  {Windus}.
\newblock {6-31G* basis set for atoms K through Zn}.
\newblock {\em The Journal of Chemical Physics}, 109(4):1223--1229, 1998.

\bibitem{Ben11}
Stefan Grimme, Andreas Hansen, Jan~Gerit Brandenburg, and Christoph Bannwarth.
\newblock Dispersion-corrected mean-field electronic structure methods.
\newblock {\em Chemical Reviews}, 116(9):5105--5154, 2016.
\newblock PMID: 27077966.

\bibitem{Ben12}
Giovanni Bussi, Davide Donadio, and Michele Parrinello.
\newblock Canonical sampling through velocity rescaling.
\newblock {\em The Journal of Chemical Physics}, 126(1):014101, 2007.

\end{thebibliography}
\bibliographystyle{unsrt}

%%%%%%%%%%%%%%%%%%%%%%%%%%%%%%%%%%%%%%%%%%%%%%%%%%%%%%%%%%%%%%%%%%%%%%%%%%%%%%%
%%%%%%%%%%%%%%%%%%%%%%%%%%%%%%%%%%%%%%%%%%%%%%%%%%%%%%%%%%%%%%%%%%%%%%%%%%%%%%%
% APPENDIX
%%%%%%%%%%%%%%%%%%%%%%%%%%%%%%%%%%%%%%%%%%%%%%%%%%%%%%%%%%%%%%%%%%%%%%%%%%%%%%%
%%%%%%%%%%%%%%%%%%%%%%%%%%%%%%%%%%%%%%%%%%%%%%%%%%%%%%%%%%%%%%%%%%%%%%%%%%%%%%%
\newpage
\appendix
\onecolumn

\section{Implementation Details}
\subsection{Details of the Base Model}
For the NequIP base, we used a latent dimension of 64 and included all irreducible representations of even and odd parity up to $l=2$.\\
We used a radial network with a radial cutoff of 4 \AA, a trainable Bessel basis of size 8 and two hidden layers of dimension 64 with SiLU nonlinearities.\\
Gated SiLU- and tanh-based nonlinearities were used for the even and odd features respectively.\\
A base with 5 interaction blocks was used.\\
The model was constructed from the official NequIP package built on top of the e3nn library \cite{e3nn}.

\subsection{Details of the Stochastic Model}
To calculate the standard deviations for the forces and energies from the invariant features produced by the NequIP base, two separate fully connected feed-forward neural networks were used.
These neural networks have an input layer of dimension 64, two hidden layers of dimension 32 and 16 respectively and an output layer of dimension 1.\\
The hidden layers have SiLU activation functions and the output layer has the exponential function as the activation function.\\
The neural network for the force standard deviations directly maps each invariant feature $\boldsymbol{v}_i$ directly to the standard deviations $\sigma_{\boldsymbol{F}_i}$.
The neural network for the energy standard deviations maps each invariant feature $\boldsymbol{v}_i$ to a scalar latent variable $u_i$ and the standard deviation is then calculated as
$\sigma_{E}=\frac{1}{n}\sum_{i=1}^n u_i$.\\
We do not normalize the force and energy labels and instead rescale the predicted means for both the forces and energies by the root mean square of the force components evaluated on the training data set. We do not rescale the standard deviations.\\
To smooth the predicted distribution for the evaluation of the log-likelihoods with more than one Monte Carlo sample, we fit a normal distribution to the mean and variance of the predicted distribution and then used the normal distribution for the evaluation of the log-likelihoods.\\

\subsection{Details of the Sampling Algorithm}
For all experiments, all models were sampled from a single Markov chain.
We always set $\alpha$ to $0.1$, $\beta$ to $0.999$ and $\epsilon$ to $10^{-3}$. \\ 
On the RMD17 dataset, a batch size of 32 was used and the learning rate $\gamma$ was exponentially decayed from $10^{-2}$ to $10^{-5}$ during the first $10^6$ training steps. After step $10^6$ the first model was sampled. Afterward, we utilized a cyclical learning rate schedule very similar to the one proposed in \cite{CMCMC}
$\gamma_i = \frac{\gamma_0}{2}\left(\cos\left(\pi+\frac{i\cdot \pi}{K}\right)+1\right)$ with $\gamma_0=0.001$ and cycle length $K=50000$ to sample the subsequent models from the same Markov chain. A single model was sampled after each cycle. The mass vector $\boldsymbol{M}$ was kept constant after the first $9\cdot 10^5$ training steps.

The same sampling procedure was used for the PEDOT dataset except a batch size of 10 was used, the initial learning rate decay and subsequent sampling occurred during the first $5 \cdot 10^5$ training steps and the mass vector $\boldsymbol{M}$ was kept constant after the first $ 10^5$ training steps.

\subsection{Details of the Dropout Neural Networks}
For the dropout architecture, a single dropout layer was added after the NequIP base in the stochastic model.\\
The resulting model was optimized using the AMSGrad optimizer with the same batch size and learning rate schedule used during the initial convergence phase of the Markov chain sampling algorithm.\\
The mean log-likelihoods were used as an objective function.\\
We used a dropout rate of 1/64 so that on average one activation was dropped per atomic energy contribution.\\
We found it necessary to use this fairly small dropout rate because higher rates significantly negatively impacted the accuracy of the resulting model, which is consistent with what was found in \cite{BNNP_DO1}.
\subsection{Evaluation of the Calibration Error and the Error Densities}
In the evaluation of the calibration errors and observed error densities, we utilize a bin width $\delta=0.01$.\\
To visualize the observed distributions of the components of the errors $(\boldsymbol{F}_i-\boldsymbol{F}_{pred\; i}), i=1,...,n$ on a test dataset, we set the observed error density as 
$\rho_{pred}(x)=\frac{f_x}{\delta}$ where $f_x$ is the fraction of errors falling into the interval $[x,x+\delta)$.

\section{The PEDOT Dataset}
The {\it ab initio} molecular dynamics on PEDOT was carried out using NWChem 6.8 \cite{Ben7}, using hybrid DFT: B3LYP \cite{Ben9,Ben8} 6-31G* \cite{Ben10} with D3(BJ) dispersion corrections \cite{Ben11} for the geometry relaxations. Previous work had indicated that diffuse functions did not significantly change the results. Damping, direct inversion in the iterative subspace (DIIS) and level shifting were turned off, and the quadratic convergence algorithm was used. The DRIVER module was used for structural relaxations, which is a quasi-newton optimization. Its default values were used, with a maximum gradient of 0.00045 and root mean square gradient of 0.00030, and a maximum Cartesian step of 0.00180 and root mean square Cartesian gradient of 0.00120. The molecular dynamics itself was carried out using the stochastic velocity rescaling (SVR) thermostat \cite{Ben12}, with a relaxation time parameter ($\tau$) of 25 fs and a timestep of 0.5 fs.

The training set is comprised of 50 randomly sampled configurations of the first 1500 timesteps from both the length 8 and length 12 PEDOTs.
Equivalently, the validation set is comprised of 15 randomly sampled configurations of the timesteps 1501-1600 from both the length 8 and length 12 PEDOTs.
The timesteps 1601-2000 served as the test sets for the respective lengths.
The test set for the length 16 PEDOT contained all 2000 configurations.
For each of the PEDOT molecules, we defined the potential energy of the ground state geometry as zero.

\section{Histogram of Force Predictions}

\begin{figure}[h!]
\begin{centering}
\onecolumn \includegraphics[scale=0.4]{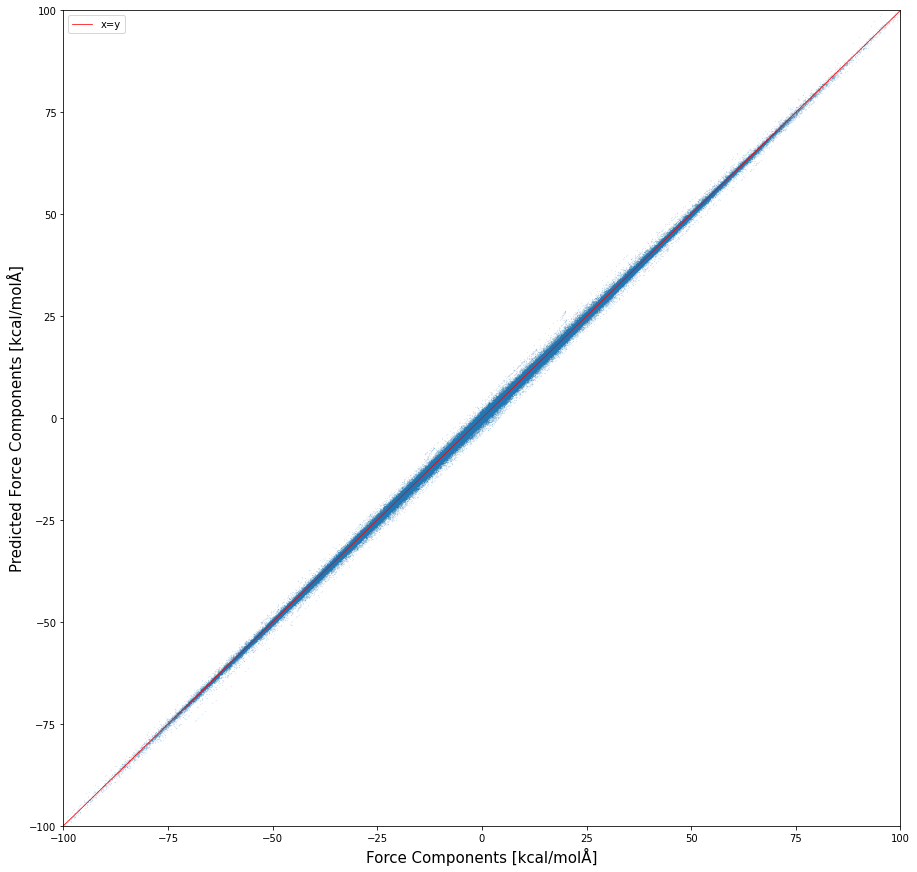}
\onecolumn \caption{Observed and predicted force components in kcal/(mol$\cdot$\AA) with k=8 Monte Carlo samples on the length 16 PEDOT dataset. }
\end{centering}
\end{figure}

\section{Additional Results on the RMD17 Dataset}

\begin{table}[h!]
%\begin{centering}
 \caption{Results of the NequIP architecture used as a base model and a single Monte Carlo sample of the Bayesian posterior of the stochastic model on the RMD17 dataset. Both models were generated with identical training, validation and test datasets.}
\centering
 \includegraphics[scale=0.85]{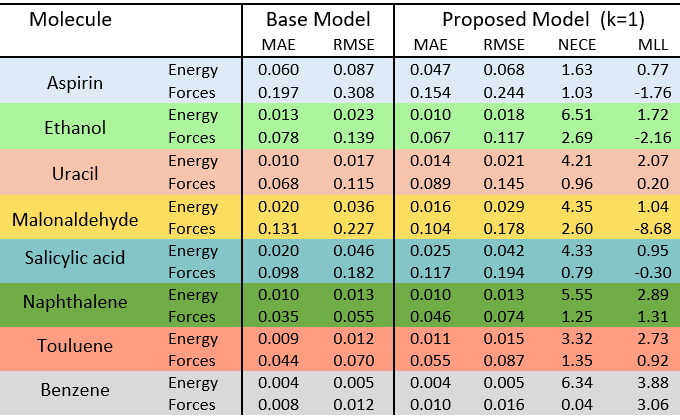}
%\end{centering}
\end{table}
%%%%%%%%%%%%%%%%%%%%%%%%%%%%%%%%%%%%%%%%%%%%%%%%%%%%%%%%%%%%%%%%%%%%%%%%%%%%%%%
%%%%%%%%%%%%%%%%%%%%%%%%%%%%%%%%%%%%%%%%%%%%%%%%%%%%%%%%%%%%%%%%%%%%%%%%%%%%%%%

\end{document}